\documentclass[aps,prl,twocolumn,groupedaddress,showpacs,amsmath,amssymb]{revtex4}
\usepackage{epsfig}
\usepackage{amssymb}
\usepackage{amsmath}
\usepackage{graphicx}
\usepackage{amsfonts}
\usepackage{epsfig}
\usepackage{revsymb}
\usepackage{bm}
\usepackage{latexsym}
\usepackage{hyperref}

\begin{document}

\title{Dynamics of the fall of a quantum $^4$He crystal  in superfluid liquid}

\author{V. L. Tsymbalenko}

\email[]{vlt49@yandex.ru}
\affiliation{NRC Kurchatov Institute, 123182 sq.Kurchatov 1 \\
                  Institute for Physical Problems RAS, 119334 Kosigin st.2 \\ Moscow, Russia}


\begin{abstract}
The motion of helium crystals has been experimentally studied when the crystals fall in the superfluid liquid owing to gravity at temperatures above the roughening transitions where the whole crystal surface is in the atomically rough state. The rate of crystal fall at $T = 1.25$K is higher than at $T = 1.54$K. This is proof of the essential role of the normal component of superfluid helium in the deceleration of crystal motion. The pressure measurements in the container have shown the effect of surface kinetics on the motions of the crystal and its size. The fall of crystals with the low surface mobility at $T=1.54$K does not change the pressure significantly. The high surface mobility at $T=1.25$K results in decreasing the pressure in the container in the course of the fall of a crystal. The pressure drop exceeds the difference in the hydrostatic pressure for the initial and final positions of the crystal. After the stop the pressure in the container relaxes to the difference mentioned above. This fact demonstrates an additional growth of the crystal in the flow of a superfluid liquid.
\end{abstract}


\maketitle

\section{Introduction}
The fast kinetics of the atomically rough surfaces of $^4$He crystals ~\cite{AP} displays a remarkable interplay between the fluid flow and the superfluid-solid interface.
This fact is mentioned by Nozieres, Uwaha ~\cite{NU}, and M.Kagan ~\cite{MK}, who theoretically studied the tangential instability of a flat liquid-solid interface.
It is found that the instability of the cylindrical crystal shape is associated with the fast interface kinetics.
The fluid flow in the direction of the cylindrical axis of a $^4$He crystal leads to developing the instability on the smaller scales  ~\cite{MaxVLT}.
As it concerns the faceted interfacial segments with low mobility, the fluid flow has no observable effect. However, there exist conditions when the crystal facets acquire high mobility comparable with that of the atomically rough surface. Provided that a large overpressure is applied to the crystalline facet, the growth rate of the facet increases drastically by 2~-~3 orders of magnitude in jump-like manner ~\cite{VLT2015}. At the initial stage the $^4$He crystal grows at high growth rates and giant accelerations of the interface. This situation is theoretically analyzed within the framework of the Rayleigh-Taylor and Richtmyer-Meshkov instabilities ~\cite{BDT}.

The experimental test of theoretical conclusions is complicated by the requirement to prepare the
fluid flow with controlled parameters while the $^4$He crystal is in the container. It has been observed that the fluid jet created with the heat flux~\cite{BKP}, the oscillating loop~\cite{VLT2013} or the motion of charges~\cite{MaxVLT} distorts the crystalline surface. Observations of the crystal shapes at the stage of fast growth demonstrate the instability due to acceleration of liquid-solid interface~\cite{BDT}. All these observations have been performed with immobile crystals under the fluid flow induced by the methods mentioned above.

Since the direct measurement of the velocity of fluid flow is difficult, the alternative method is to set a $^4$He crystal into motion in the immobile fluid. The velocity of the crystal can be directly measured using video recording of the crystal motion. The force exerted to the crystal is determined by the method of setting the crystal into motion. In a series of experiments a superconducting loop is used for forcing the crystal to oscillate. The crystal is either pierced by the crossbar or is fixed on the crossbar~\cite{VLT2013}. The oscillation frequency of such system is determined by the crystal mass and force arising from the fluid flow. The maximum velocity of the crystal, which one can reach, is as large as 3 cm/s. In the first case, a further enhancement of the oscillation amplitude results in the fast remelting of the crystal before its detaching from the crossbar. In the second case the crystal escapes from the crossbar.

Another method is the study of free fall of a $^4$He crystal in the field of gravity. Here the motive force is the weight of the crystal, compensated partially by the force of buoyancy. A direct observation of the crystal shape transformation during the fall of the crystal has been performed by the Japanese group at 0.3K using a high-speed camera~\cite{NOTY}. At these temperatures the shape of the $^4$He crystal is governed by the facets with low growth rates. As can be seen from this study, the effect of facet interface kinetics upon the motion of the $^4$He crystal is insignificant.

In experiments which continue the study of burst-like growth effects within the 0.1~--~0.2K range we have noticed a curious feature~\cite{VLT2019}. As the crystal stops to grow, the mobility of crystalline facets remains very high~\cite{VLT2004}. The crystal starts to remelt in the hydrostatic pressure gradient and then detaches from its nucleation site at the wall in the upper part of the container. The fall of the crystal is accompanied by a decreasing pressure in the container. It is unexpected and intriguing that the pressure changes non-monotonically. At first, the pressure drops below the difference in hydrostatic pressures between the crystal nucleation site and the container bottom. Then the pressure relaxes to the pressure difference indicated. Such behavior of the pressure implies two essential facts. Firstly, the pressure decrease in the course of fall is possible only provided that the crystal is growing and its volume increases. In other words, as the crystal stops to grow in the burst-like growth regime, the crystal facets keep their high mobility. Secondly, the extra pressure drop and the following pressure relaxation prove unambiguously the increase of the crystal volume during the fall of the crystal not only due to hydrostatic pressure but also due to an additional factor.

In Ref.~\cite{VLT2019} the influence of fluid flow is involved as a factor explaining such an effect. The simplified model is suggested for the fall of an isotropic sphere with a mobile interfacial boundary which is encircled by the fluid flow. The fluid flow is supposed to be laminar as in all theoretical works known so far~\cite{NU, MK, MaxVLT, BDT}. The model has explained the pressure behavior and given the estimate for the magnitude of the averaged kinetic growth coefficient $K=0.22$~s/m, $V = K \Delta \mu$. Here $V$ is the interface growth rate and $\Delta \mu$ is the difference in chemical potentials. However, the time behavior of pressure $p(t)$ measured experimentally in the course of the fall differs noticeably from the prediction of the model. The reasons may be as follows: growth anisotropy of crystal facets, specific trajectory of the fall and orientation of the crystal, specific features of fluid flow at such velocities of encircling and so on.

Many of these difficulties can be avoided by observing the fall of helium crystals at 1.2~-~1.6K. Within this temperature range all the liquid-solid $^4$He interface is in the atomically rough state. The kinetic growth coefficient is isotropic. The crystals have approximately spherical shape~\cite{VLT1995}. The numerical magnitude of the kinetic growth coefficient for such temperatures lies within the range $K \sim 0.05 - 0.3$~s/cm, i.e. it is of the order of the magnitude given in Ref.~\cite{VLT2019}. Comparing the dynamics of the fall of a crystal in the superfluid with different normal and superfluid densities, we can clarify and understand the role and contribution of normal and superfluid components to the dynamics of motion of $^4$He crystals.

\section{Experimental methods and results}

The crystals are grown in optical container similar to those used in experiments previously ~\cite{VLT1996,VLT1997}. The inner diameter of the container is 20~mm and the volume is 10~cm$^3$.
A capacitive pressure sensor is fixed on the upper flange of the container. The pressure is measured at 2~ms time intervals. The crystal nucleates at a tungsten tip located at 10.5~mm distance from the bottom of the container. A high voltage is applied to the tip. After nucleation and growth, the crystal has 1-2 mm diameter. The hydrostatic pressure gradient leads to a remelting of the crystal, while its center shifts down and it subsequently separates from the tip~\cite{VLT1995}. The fall of the crystal is recorded by a television system with a ccd-matrix. The crystal is illuminated by a light pulse of an infrared LED lasting 5~$\mu s$  and synchronized with the frame. An example of shooting the fall of a crystal at two temperatures is shown in Fig.1.
\begin{figure}
\begin{center}
\includegraphics[%
  width=0.65\linewidth,
  keepaspectratio]{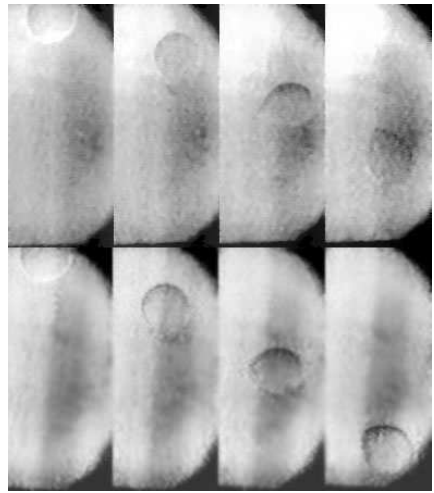}
\end{center}
\caption{The crystal fall at two temperatures. The upper series is made at T = 1.54K, and the lower series at T = 1.25K. The intervals between the frames from left to right: 60ms, 40ms and 40ms. The light pulse duration is 5 $\mu s$.}
\end{figure}
The relaxation time of the pressure in the container to the pressure of the external system is $\sim$4~s, i.e., the crystals fall over an interval $\sim$160~ms at almost constant mass of helium in the container. Note that when the crystal is separated from the tip, a protrusion remains on its surface, as was previously observed in Ref.~\cite{VLT1995}. Then, the protrusion flattens and at lower temperature this occurs faster due to the higher kinetic growth coefficient. The observation for the protrusion reveals that the crystal rotates in the course of its fall.

The experiments were carried out at two temperatures. At T=1.54K, the crystal has the bcc structure, and at T=1.25K the crystal has the hcp structure. The kinetic growth coefficient of an atomically rough surface in this temperature range is determined by the energy dissipation processes in the liquid. These processes are insensitive to the crystal structure~\cite{MaxVLT2}. The surface growth rate in both cases is linearly dependent on supersaturation and isotropic for small deviations from equilibrium. The kinetic growth coefficients measured from surface relaxation~\cite{BLN} and from pressure relaxation~\cite{MaxVLT2} are consistent and equal to $K = 0.063 \pm 0.012 m/s$ at 1.54K and $K = 0.23 \pm 0.03 m/s$ at 1.25K.

Fig.2 shows the results of measurements on the position of the crystal center $H$ as a function of the fall time. It can be seen that the temperature growth decelerates during the fall.
\begin{figure}
\begin{center}
\includegraphics[%
  width=0.65\linewidth,
  keepaspectratio]{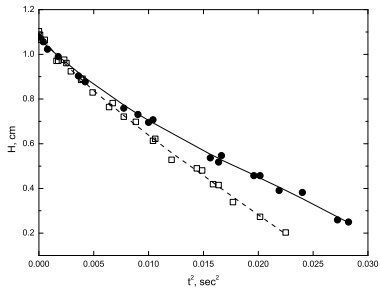}
\end{center}
\caption{The position of crystal center versus the square of the fall time. The ordinate corresponds to the distance from the center of the crystals to the bottom of the container. The squares refer to the crystals grown at 1.25K and circles at 1.54K}
\end{figure}
This can also be seen in the video of Fig.1, where the crystals starting from the same position are at different heights after 140~ms. In Fig.3, top panel, we see the time behavior of the fall rate $V(t)$ of crystals, obtained by differentiating the smoothed data $H(t)$.
\begin{figure}
\begin{center}
\includegraphics[%
  width=0.65\linewidth,
  keepaspectratio]{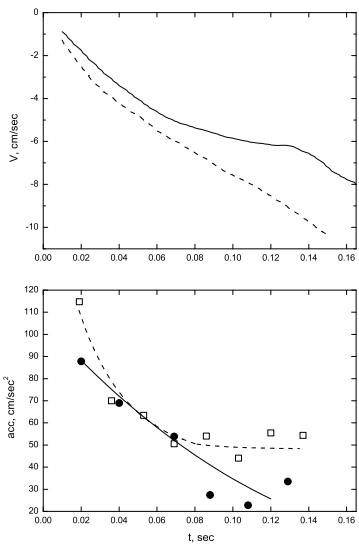}
\end{center}
\caption{Above: The time behavior of the fall rate of crystals showing the velocity $V(t)$ as a
function of time. Below: the change in the acceleration of fall over time. The squares refer
to the crystals grown at 1.25K and circles at 1.54K The solid curves refer to 1.54K and the
dashed lines to 1.25K.}
\end{figure}
Crystals fall with acceleration. Lowering the temperature increases the fall rate of crystals. The
wave on the dependence $V(t)$ at T~=~1.54K is probably caused by the rotation of the crystal and the non-sphericity of its shape. The Fig. 3, bottom panel, shows the change in acceleration in the process of fall, obtained by numerical differentiation of the upper graph. Acceleration is determined by the points included within the 20~ms time interval. To avoid dispersion in the velocity graph at T~=~1.52K, the time interval is enlarged to $\pm 30$~ms for the point at 130~ms. The low accuracy in determining the acceleration allows us to draw qualitative conclusions only. The acceleration of the crystal motion decreases with time at both temperatures. The magnitudes of acceleration are within the range of $30-100 cm / s^2$. It is possible that the acceleration at $\sim 1.25 K$ becomes constant after $\sim 60 ms$.

Averaged over all measurements the results for the pressure variations are shown in Fig.4.  To reduce noise, the records of $\Delta p (t)$ in each series are combined by the time of separating the crystal from the tip and summarized.  One can see that the pressure variation differs drastically for two temperatures.
\begin{figure}
\begin{center}
\includegraphics[%
  width=0.65\linewidth,
  keepaspectratio]{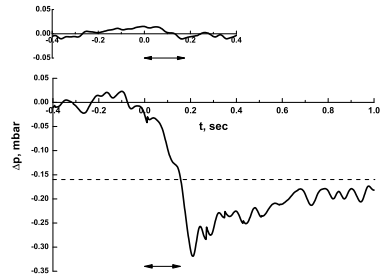}
\end{center}
\caption{Upper curve: the pressure change in the container during the fall of the crystal at
1.54K. The pressure difference between the beginning and the end of the fall is 0.026mbar.
This value lies within the measurement noise magnitude of 0.03mbar. Bottom curve: $\Delta p(t)$
during the fall of the crystal at 1.25K. The arrows show the time of crystal fall. The dashed
line corresponds to the difference in the hydrostatic pressure between the initial and final
positions of the crystals.}
\end{figure}
At high temperature, the change in pressure during the fall, if it occurred, does not exceed the measurement noise of $\sim 0.03$mbar. At low temperatures, a pressure drop is observed similar to that observed at 106~mK for crystals whose surface retains mobility after nucleation in the burst-like growth mode ~\cite{VLT2019}. As in that case, the pressure drops below the difference in the hydrostatic pressure between the place of crystal nucleation and the bottom of the container. Then, pressure relaxation to this magnitude is observed, which in our case is $\Delta p_0 = 0.16$ mbar. The same kinetic growth coefficients for the crystals that grow at different temperatures in the completely different states lead to a qualitatively identical change in pressure upon falling.

The observations on the fall of the crystals at 1.16K show that the crystals have their
flat facets facing down, resulting in floating during the motion, see Fig.5. The lower facet is oriented perpendicular to the direction of fall. The moment of forces acting on the body from the side by the liquid flow turns the crystal in this way~\cite{LAMB}.
\begin{figure}
\begin{center}
\includegraphics[%
  width=0.65\linewidth,
  keepaspectratio]{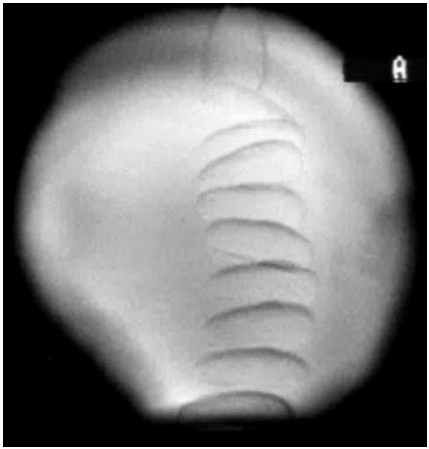}
\end{center}
\caption{The crystal fall illuminated with stroboscopic light: pulse duration 5 $\mu$s, pulse interval 20 ms, 1.16K}
\end{figure}
Below 0.6 K, the crystal facets have such low growth rate that the nucleated crystals remain hanging on the tip for tens of minutes, as was previously observed~\cite{VLT1995}.

\section{Discussion of the results}
The crystal falls at temperatures where the normal component density of superfluid helium is significant. Since the rates of fall are small and the fluid can be considered as incompressible, the equations of hydrodynamics are separated into the equations for the superfluid and normal components~\cite{LL}. The equation for the normal component reads
\begin{equation}\label{eq01}
\frac{\partial{\vec v_n}}{\partial t}  +(\vec v_n \nabla)\vec v_n= -\frac{1}{\rho_n}\nabla p_n + \frac{\eta_n}{\rho_n}\triangle \vec v_n
\end{equation}
Here $\vec{v_n}$, $p_n$, $\eta_n$ and $\rho_n$ are the velocity,  pressure, viscosity  and density of the normal component, respectively. The equation can be transformed to the dimensionless representation similar to the classical Navier-Stokes equation:
 \begin{eqnarray}\label{eq02}
t \rightarrow t \frac{R}{U}, r \rightarrow r R, v_n \rightarrow v_n U,  p_n \rightarrow \rho_n U^2 p_n
\nonumber \\
\frac{\partial{v_n}}{\partial t}  +(v_n \nabla) v_n= -\nabla p_n + \frac{1}{Re_n}\triangle \vec v_n,  \\
Re_n =UR\frac{\rho_n}{\eta_n} =\frac{UR}{\nu_n}. \nonumber
\end{eqnarray}
Here $U$ is the velocity of the body, $R$ is its radius and $Re_n$ is the Reynolds parameter for the normal component flow. The normal component density is $0.41\rho$ at 1.54K, $0.14\rho$ at 1.25K  and $0.08\rho$ at 1.16K~\cite{Maynard}. The viscosity of the normal component  is 20 $\mu P$ at 1.54K and  23 $\mu P$ at 1.25K ~\cite{EHH}. For the crystal radius of $\sim 1 mm$ and falling rates of 1-10 cm/sec, the Reynolds parameter varies within Re$_n$ = 400 -- 4000 at 1.54K and Re$_n$ = 100 -- 1000 at 1.25K. There exist a lot of studies on the motion of spheres in helium at saturated vapor pressure. Without attempting to provide a complete description of these results, we use only a few works for illustrative purposes. The resistance to the motion of a ball in normal helium is proportional to velocity squared and estimated by the relation ~\cite{BHH,DHH,LR}
 \begin{equation}\label{eq03}
F_n = Cx_n \frac{1}{2}\rho_n U^2 \pi R^2.
\end{equation}
The parameter $Cx_n$ is 0.2-0.5 in the indicated range of the Reynolds numbers.

The superfluid component flow is described by the ideal fluid equation
 \begin{equation}\label{eq04}
\triangle \varphi = 0, \vec v_s = \nabla \varphi.
\end{equation}
The experimental studies of the motion of spherical bodies in superfluid helium show  that resistance to motion is observed starting at velocities of $\sim 0.1$ cm/s~\cite{A}. Feynman explained this phenomenon by the production of vortices during helium flow around an object and gave an estimate for the critical velocity of such process, $v_{crit} = (h/m_4 R) ln(R/A)$  ~\cite{F}, where $h$ is the Plank constant, $m_4$ is the helium atom mass and $A$ is the interatomic distance. The resistance to the motion of the sphere is approximately described by a relation similar to expression (\ref{eq03})~\cite{A,Vinen,Sch}
 \begin{equation}\label{eq05}
F_s = Cx_s \frac{1}{2}\rho_s U^2 \pi R^2.
\end{equation}
The parameter $Cx_s$ in the indicated range of the Reynolds numbers is 0.3-0.5. A number of authors use the Reynolds parameter $Re_s= Rv_s/ \nu_s$ to evaluate the nature of superfluid flow and turbulence. The value $\nu_s~ = h / m_4 ~\simeq~ 10^{-3}$ cm$^2$/s  is taken as the kinematic viscosity. For the rates of crystal fall in these experiments, this corresponds to the values $Re_s = 100-1000$.

The equations (\ref{eq02}) and (\ref{eq04}) are related via the boundary condition on the crystal surface. Andreev and Knizhnik ~\cite{AK}  proposed the condition $v_{surf} =v_n$, that is, the mass flow through the phase boundary is provided only by the superfluid component. This statement is consistent with the experimental fact that near $\lambda$-transition, when $\rho_s \rightarrow 0$, the kinetic growth coefficient also tends to zero ~\cite{MaxVLT2}. Therefore, the motion of a sphere in the normal component (\ref{eq02}) occurs as the motion of an impenetrable ball, whose size and shape, generally speaking, change during the motion.  The superfluid flow around a sphere with the mass flow through its surface is considered in Ref.~\cite{VLT2019} under the assumption that the superfluid flow is laminar. In this case, the crystal grows mainly in the direction perpendicular to the flow of the liquid. This change in the shape leads to increasing  the added mass and, as a result, to decreasing the acceleration of movement. In our experiments, as is seen in Fig.1, the crystals rotate during the fall. This leads to more uniform surface growth and small change in the shape of the crystal. It follows that the added mass of the superfluid liquid also changes insignificantly, and does not  lead to reducing the acceleration during a fall within the framework of the model in Ref.~\cite{VLT2019}.

In our experiments, the flow around a crystal differs significantly from that considered in Ref.~\cite{VLT2019}. The rotation of the crystal leads to the appearance of a Magnus force perpendicular to the direction of motion. The formation of vortices in the normal and superfluid components creates a complex picture of the pressure distribution on the crystal surface. For these reasons, the flow of liquid around the crystal is very complicated. The problem of crystal growth
taking into account these factors has not yet been solved.

For the values of the Reynolds parameter in these experiments (see above), the crystal falls down with the formation of a vortex pattern in the normal component and the creation of quantum vortices in the superfluid component. Combining formulas (\ref{eq03}) and (\ref{eq05}), we obtain an expression for the total force $F$ acting on the crystal
 \begin{equation}\label{eq06}
F = (Cx_n \kappa_n+Cx_s \kappa_s)\frac{1}{2}\rho U^2 \pi R^2, \kappa_n+\kappa_s=1.
\end{equation}
Here the coefficients $\kappa_s$ and $\kappa_n$ are the fractions of the superfluid and normal components, respectively. The equation of motion of the crystal reads
 \begin{equation}\label{eq07}
 \rho' \alpha\frac{4 \pi}{3}R^3 \dot{U}= -(Cx_n \kappa_n+Cx_s \kappa_s)\frac{1}{2}\rho U^2 \pi R^2 + \Delta\rho \frac{4 \pi}{3}R^3g ,
\end{equation}
where  $\rho'$ is the density of the crystal,  $\Delta\rho$ is the difference in density between the solid and liquid phases, and  $\alpha$ takes the added mass of the liquid into account. For a sphere, the parameter $\alpha$ is equal to $\alpha=1+\frac{\rho}{2\rho'} \simeq 1.45$. The solution of equation (\ref{eq07}) has the form
\begin{eqnarray}\label{eq08}
U(t)=U_0 \tanh\left(\frac{t}{\tau}\right),  <Cx> = Cx_n \kappa_n+Cx_s \kappa_s, \\
U_0=\left(\frac{8}{3}  \frac{\Delta\rho}{\rho}  \frac{gR}{<Cx>}\right)^{\frac{1}{2}}
\simeq 0.54\sqrt{\frac{gR}{<Cx>}},\nonumber \\
\tau =\alpha \left(\frac{8}{3}  \frac{(\rho')^2}{\rho\Delta\rho}  \frac{1}{<Cx>} \frac{R}{g}\right)^{\frac{1}{2}}
\simeq 7.93\sqrt{\frac{R}{g<Cx>} } . \nonumber
\end{eqnarray}
From the form of equation (\ref{eq07}) it follows that if the condition $Cx_n >Cx_s$ is fulfilled, then the lower rate of crystal fall at higher temperatures is explained by  increasing the normal component fraction. The values of the $Cx_{n,s}$ coefficients at a liquid pressure 25~bar are unknown. Assuming these values to be of the order of $Cx_{n,s}$ at the saturated vapor pressure for solid balls ~\cite{A}, we obtain an estimate for the parameters $U_0$ and $\tau$. For $<Cx>= 0.5$, $U_0 \approx 8$ cm/s and $\tau \approx 120$~ms. For $<Cx>= 0.1$, $U_0 \approx 17$~cm/s and $\tau \approx 260$~ms. As can be seen from Fig. 2, the second pair of parameters is better consistent with  the experimental data. These estimates should be considered as qualitative since the essential features of the process, for example, such as the rotation of the crystal and the possible dependence of the Reynolds parameter on velocity, are not taken into account when deriving equation (\ref{eq07}).

The decrease in pressure during the fall of the crystal is clear evidence for the growth of its volume. The change in crystal volume is due to two factors. First, due to growth in the denser liquid layers at the bottom of the container, where the pressure exceeds that for the liquid in the initial position of the crystal by $\Delta p_0 = \rho g \Delta h = 0.16$mbar. The second source of the bulk growth is the effect of liquid flow. The qualitative change in the shape of a crystal is clear from the general picture of the liquid flow around the crystal. The flow rate of the liquid is maximum for the cross section which gives the maximum difference in chemical potentials and the maximum growth rate~\cite{VLT2019}. The crystal starts to increase in transverse dimension and volume. We estimate its value by the magnitude of pressure relaxation after the motion stops, $\Delta p_{rel} \simeq 0.15 mbar$. The total increase in the crystal volume with an initial size of $\sim 1 mm$ will be $\sim 30 \%$. The transverse crystal size will increase by about  0.2~mm. The video recording has a resolution of 0.04~mm~per~pixel, which prevents from measuring this effect with the necessary accuracy. Owing to the rotation of the crystals, it is also impossible to check carefully the model ~\cite{VLT2019} predicting an increase in the crystal size mainly in the direction perpendicular to the liquid. A comparison of the measurements of $p(t)$ at two temperatures shows, see Fig. 4, that a decrease in the fraction of the superfluid component leads to a practically constant volume of the crystal during the fall. This results from the small pressure variation in the container. As is noted above, this conclusion agrees with the observation that the crystal grows from the superfluid component.

\section{Conclusion}
The study of falling the crystals in superfluid helium has shown a significant effect of the normal component on the resistance to motion. This is clearly seen in the dependencies $V(t)$ in Fig.3.
For  the fluid of the 41\% normal component fraction, the velocity of motion is noticeably lower
 than that in a fluid with the 14\% normal component concentration. The influence of the interface growth kinetics is clearly shown by the pressure measurements during the fall. The low kinetic coefficient of interface growth at 1.54K does not result in a noticeable crystal growth either due to gradient of hydrostatic pressure or pressure of the liquid flow. The increase in the interface mobility at 1.25K demonstrates these both effects. When falling, the pressure decreases due to crystal growth under influence of both effects, and then increases after stopping the crystal at the bottom of the container due to melting the excess crystal volume as a result of the liquid flow effect. The experimental results have confirmed the hypothesis that the nonmonotonic behavior pressure observed for the fall of crystals nucleated in the burst-like growth mode at $\sim 0.1$~K is associated with the remaining high kinetics of crystal facets after their growth. The growth rate of crystal facets for $\sim 0.6$s after growing a crystal in the burst-like growth mode \cite{VLT2019} exceeds the growth rate of crystal facets in the normal state by $\sim20$~ times \cite{Balibar}.

\begin{acknowledgements}
The author is grateful to V.~V.~Dmitriev for the possibility of performing these experiments at Kapitza Institute for Physical Problems RAS. The author is also grateful to V.~V.~Zavyalov for supporting this work, S.N.Burmistrov for helpful comments and V.~S.~Kruglov for interest to the work.
\end{acknowledgements}


\begin{thebibliography}{99}
\bibitem{AP}	A.~F.~Andreev and A.~Ya.~Parshin, {\it Sov. Phys. JETP} {\bf 48}, 763 (1978).
\bibitem {NU} P.~Nozieres and M.~Uwaha, {\it J.de Physique}, {\bf 47}, 263 (1986)
\bibitem{MK}	M.~Yu.~Kagan, {\it JETP} {\bf 63}, 288 (1986).
\bibitem{MaxVLT} L.~A.~Maksimov and  V.~L.~Tsymbalenko,  {\it JETP} {\bf 95}, 455 (2002).
\bibitem{VLT2015} V.~L.~ Tsymbalenko, {\it Physics-Uspekhi}, {\bf 58}, 1059 (2015)
\bibitem {BDT} S.~N.~Burmistrov, L.~B.~Dubovskii and V.~L.~Tsymbalenko,, {\it Phys.Rev.E}, {\bf 79}, 051606 (2009)
\bibitem{BKP}	A.~V.~Babkin, D.~B.~Kopeliovich and A.~Ya.~Parshin, {\it Sov. Phys. JETP} {\bf 62}, 1322 (1985).
\bibitem{VLT2013} V.~L.~Tsymbalenko,  {\it J. Low Temp. Phys.} {\bf 171}, 21 (2013)
\bibitem {NOTY}  R.~Nomura, Y.~Okuda, A.~Tachiki and T.~Yoshida,  {\it New J.Phys}., {\bf 16}, 113022 (2014)
\bibitem {VLT2019}  V.~L.~Tsymbalenko,  {\it J.Low Temp.Phys.}, {\bf 195}, 153 (2019)
\bibitem {VLT2004} { V.~L.~Tsymbalenko,} {\it JETP}, {\bf 99}, 1214 (2004)
\bibitem {VLT1995} { V.~L.~Tsymbalenko,} {\it Ukr.Low Temp.Phys.}, {\bf 21}, 120 (1995)
\bibitem {VLT1996} { V.~L.~Tsymbalenko,} {\it Cryogenics}, {\bf 36}, 65 (1996)
\bibitem {VLT1997} { V.~L.~Tsymbalenko,} {\it Instr.Exp.Tech.}, {\bf 40}, 585 (1997)
\bibitem {MaxVLT2} { L.~A.~Maksimov and V.~L.~Tsymbalenko,} {\it JETP}, {\bf 87}, 714 (1998)
\bibitem {BLN} { J.~Bodensohn, P.~Leiderer and K.~Nicolai,} {\it Z.Phys.B}, {\bf 64}, 55 (1986)
\bibitem{LAMB} H.~Lamb,  {\it Hydrodynamics}, (University Press, Cambridge, 1916).
\bibitem{LL}	L.~D.~Landau and E.~M.~Lifshits, \textit{Fluid Mechanics}, (Pergamon Press, 1987).
\bibitem {Maynard} { J.~D.~Maynard,} {\it Phys.Rev.B}, {\bf 14}, 3868 (1976)
\bibitem {EHH} { K.~M.~Eisele and A.~C.~Hollis Hallett,} {\it Can.J.Phys}, {\bf 36}, 25 (1958)
\bibitem {BHH} { C.~B.~Benson and A.~C.~Hollis Hallett,} {\it Can.J.Phys.}, {\bf 34}, 668 (1956)
\bibitem {DHH} { R.~J.~Donnelly and  A.~C.~Hollis Hallett,} {\it Ann.Phys.}, {\bf 3}, 320 (1958)
\bibitem {LR} { R.~A.~Laing and H.~E.~Rorschach, Jr.,} {\it Phys.Fluids}, {\bf 4}, 564 (1961)
\bibitem {A} { J.~P.~Adelin, Jr.03,} {\it https://thesis.library.caltech.edu/2383/} (1967)
\bibitem {F} { R.~P.~Feynman,} \textit{Progr.In Low Temp.Phys}, {\bf 1}, (Amsterdam, 1955)
\bibitem {Vinen} { J.~J.~Niemela and W.~F.~Vinen,} {\it J.Low Temp.Phys.}, {\bf 128}, 167 (2002)
\bibitem {Sch} { W.~Schoepe,} {\it JETP Lett.}, {\bf 102}, 117 (2015)
\bibitem {AK} { A.~F.~Andreev and V.~G.~Knizhnik,} {\it Sov.Phys.JETP}, {\bf 83}, 416 (1982)
\bibitem {Balibar} { S.~Balibar, F.~Gallet, P.~Nozieres, E.~Rolley, P.~E.~Wolf,} {\it J.de Physique}, {\bf 46}, 1987 (1985)
\end{thebibliography}
\end{document}